\documentclass[aps,prb,twocolumn,showpacs]{revtex4}

\usepackage[dvips]{graphicx}
\usepackage{psfrag}
\usepackage{amsmath}
\usepackage{amssymb}
\def \bm{\boldsymbol}
\begin{document}

\title{Study of structure and lattice dynamics of the
Sr$_2$CuO$_2$Cl$_2$(001) surface\\ 
by helium-atom scattering}
\author{M. Farzaneh}
\author{XF. Liu}
\author{M. El-Batanouny}
\affiliation{Department of Physics, Boston University, 590
  Commonwealth Avenue, Boston Massachusetts 02215}
\author{F.C. Chou}
\affiliation{Center for Materials Science and Engineering, Massachusetts
  Institute of Technology, Cambridge, Massachusetts 02139}

\begin{abstract} 
Structure and lattice dynamics of the (001) surface of Sr$_2$CuO$_2$Cl$_2$
have been studied by helium atom scattering (HAS). Analysis of diffraction
patterns obtained by elastic HAS  revealed a surface periodicity consistent
with bulk termination,  and confirms that the surface is non-polar and
stable which favors a SrCl surface termination.  Bulk and surface lattice
dynamical calculations based on the shell-model were carried out to
characterize the experimental phonon dispersions obtained by inelastic
HAS. No experimental surface mode was observed above 200 cm$^{-1}$.
Comparison between the experimental data and theoretical results for two
different slabs with SrCl and CuO$_2$ terminations showed that the
experimental data conforms exclusively with the SrCl surface modes.
\end{abstract}
%\pacs{}
\maketitle

\section{INTRODUCTION}
Surface lattice dynamics of metals, semiconductors, simple insulators
and layered structures \cite{surf-phonons} have been studied
extensively . However, high temperature superconductors (HTSC), their
parent compounds and related lamellar cuprates are more complex in
structure and dynamics and the study of their surface dynamical
properties has been limited to only a few cases
\cite{Paltzer96,Paltzer98}. Because of the established relation
between phonons and superconductivity in conventional superconductors,
bulk phonons of the HTSC and their parent compounds received
considerable attention in the early years following their
discovery. Optical spectroscopies (infrared (IR) and Raman)
\cite{Mostoller90,Eklund89,Crawford89,Burns88,Weber88,Ohana89} and
neutron scattering experiments
\cite{Pintschovius91,Boni88,Birgeneau87} were used to study the
lattice dynamics of these compounds. Although it is believed that spin
interactions play an important role in the superconducting properties
of HTSC cuprates, the role of the electron-phonon coupling still
remains a puzzle.  For example, the isotope effect in the optimally
doped cuprates is very small \cite{Hoen89}, but away from optimal
doping a large isotope effect has been observed \cite{Crawford90}. In
addition, the highest longitudinal optical (LO) phonon branch softens
and shows line-width broadening when the parent compounds of several
HTSC, such as La$_{2-x}$Sr$_x$CuO$_4$ \cite{McQueeney99,Pintschovius99}
and YBa$_2$Cu$_3$O$_{6+y}$ \cite{Reichardt89}, are hole-doped to a
superconducting state.  Furthermore, in recent angle resolved
photoemission spectroscopy (ARPES) experiments on several HTSC
materials, an abrupt change of electron velocity at 50-80 meV has been
observed \cite{Lanzara01} which is interpreted to be caused by the
coupling of electrons to the LO phonon modes in the same energy range.
Whether these effects suggest that electron-phonon coupling in HTSC
can drive electron-pairing or it is just a secondary mechanism
that occurs naturally in an interacting system without affecting the
superconducting pairing mechanism is still open to discussion and
interpretation \cite{Sandvik04}.  Therefore the role of phonons in
HTSC is still not completely understood and electron-phonon coupling
can still effect the superconducting state. For this
reason lattice dynamical studies of these compounds and related
cuprates are still important.

\begin{figure}
\begin{center}
  \includegraphics*[width=3in]{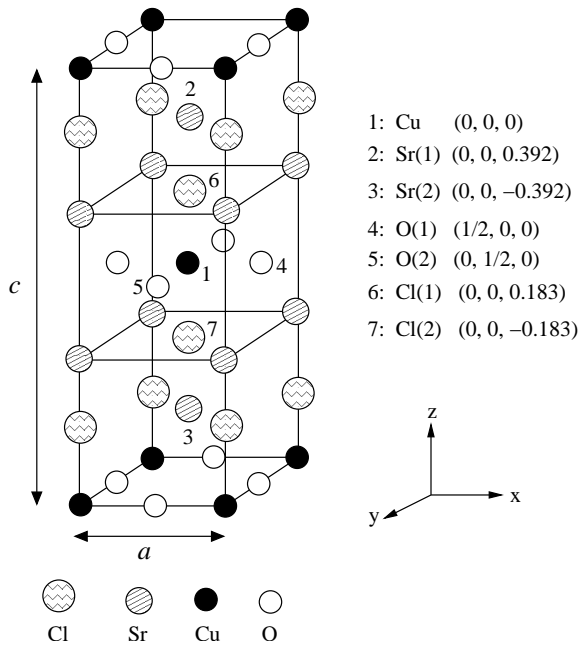}
  \caption{Crystal structure and positions of the atoms in
    Sr$_2$CuO$_2$Cl$_2$. Atomic positions are taken from 
    Ref. \onlinecite{Miller90}.}
\label{crystal}
\end{center}
\end{figure}

In this paper we report on the study of the structure and dynamics of the
(001) surface of Sr$_2$CuO$_2$Cl$_2$.  In the bulk, this compound is a
layered perovskite with the body centered tetragonal structure of
K$_2$NiF$_4$ type and space group $I4/mmm(D_{4h}^{17})$\cite{Miller90}. It
consists of CuO$_2$ sheets separated by double layers of SrCl. Neutron
diffraction studies\cite{Miller90} show that the lattice parameters for
Sr$_2$CuO$_2$Cl$_2$ at room temperature are $a=3.9716$ {\AA} and $c=15.6126$
{\AA}. The crystal structure of Sr$_2$CuO$_2$Cl$_2$ is shown in Figure
\ref{crystal}.
Sr$_2$CuO$_2$Cl$_2$ is also isostructural to the high-temperature phase of
La$_2$CuO$_4$, but with the out of plane oxygen ions at the apices of the
CuO$_6$ octahedra replaced by Cl and the La by Sr.

However two clear differences between Sr$_2$CuO$_2$Cl$_2$ and
La$_2$CuO$_4$ stand out: unlike La$_2$CuO$_4$, which undergoes a structural phase
transition from tetragonal to orthorhombic symmetry at about 530 K
\cite{Kastner98} involving rotations of the CuO$_6$
octahedra, no evidence of a structural distortion has been observed in
Sr$_2$CuO$_2$Cl$_2$ down to 10 K \cite{Miller90,Vaknin90}. Also in contrast
to La$_2$CuO$_4$, which can be doped by substitution of La with a
divalent ion like Sr$^{+2}$ and shows a superconducting behavior below
$T_c\sim 40$K, neither electron nor hole doping of Sr$_2$CuO$_2$Cl$_2$
has been possible\cite{Hiroi96}, although the similarly structured
Ca$_2$CuO$_2$Cl$_2$ can be doped successfully by Na with
$T_c\sim$ 28 K \cite{Ohishi04} and very recently superconductivity 
at $T_c\sim$ 30 K has been reported in apical oxygen doped
Sr$_2$CuO$_{2+\delta}$Cl$_{2-y}$ \cite{QQLiu05}.    

Neutron diffraction experiments
\cite{Vaknin97,Wang90,Vaknin90,Kim01,Greven94} have been used to
explore the similarities and differences between the magnetic
properties of Sr$_2$CuO$_2$Cl$_2$ and La$_2$CuO$_4$ and to understand
the magnetic interactions in these two-dimensional (2D) spin-1/2
systems. In particular, it is shown that both Sr$_2$CuO$_2$Cl$_2$ and
La$_2$CuO$_4$ are antiferromagnetic (AFM) insulators with N\'{e}el
temperatures ($T_N$) of about 250 K and 300 K, respectively, and have
similar intralayer exchange coupling $J$ of about 130 meV; in the
paramagnetic phase both are good examples of 2D square lattice quantum
Heisenberg antiferromagnets (2DSLQHA) \cite{Greven94,Kastner98}.  The
main difference between the magnetic properties of these two compounds
is that the orthorhombic distortion in La$_2$CuO$_4$ allows for a
Dzyaloshinsky-Moriya (DM) antisymmetric exchange interaction in the
spin Hamiltonian which is absent in Sr$_2$CuO$_2$Cl$_2$ because of the
stability of its tetragonal structure to the lowest temperatures
measured \cite{Kastner98}. The DM interaction in La$_2$CuO$_4$ leads
to a weak ferromagnetic moment perpendicular to CuO$_2$ planes
\cite{Thio88}. However, the weak inter-layer coupling causes the
moments of successive layers to order antiferromagnetically, so the
weak ferromagnetism is hidden in the N\'{e}el state at zero field
\cite{Thio94,Kastner98}.

In light of the above discussions, and the fact that
Sr$_2$CuO$_2$Cl$_2$ has the advantage that it can be easily cleaved, a
study of the surface lattice dynamics of Sr$_2$CuO$_2$Cl$_2$ can not
only provide information about the bulk phonons of this compound,
which are currently limited to a few optical spectroscopy studies, but
also provide insight as to the modifications effected at the
surface to the vibrational modes of the bulk for La$_2$CuO$_4$, where
the bulk dynamics has already been studied but no surface phonon study
exists.

\section{EXPERIMENTAL PROCEDURE}
The single crystalline samples of Sr$_2$CuO$_2$Cl$_2$ used in our experiments
were grown at MIT crystal growth facility. The samples were plate like with
the $c$-axis perpendicular to the sheets of the plate. They were cut to
dimensions of about 7$\times$7$\times$0.5 mm and mounted on a copper sample
holder by silver conducting epoxy. A cleaving post was then attached to the
sample surface, also by the epoxy.  The prepared sample was mounted on a
manipulator which accommodated xyz motions, as well as polar and azimuthal
rotations.  The {\em in situ} ultra-high vacuum (UHV) cleaving of the sample
was implemented by the top-post method by knocking off the post on the
surface. Immediately after cleaving, the quality of the long range ordering
on the surface was checked by low energy electron diffraction (LEED). The
LEED images were sharp spots and ordered in a square pattern, with no
discernable satellites.

Elastic and inelastic He-atom scattering (HAS) techniques were used
for measuring the structure and dynamics of the surface.  Since about
1975, novel surface spectroscopic methods based on the interaction of
thermal energy neutral helium atoms with solid surfaces have become
possible due to the progress in combining high-resolution He beam
production systems with UHV techniques. Because of their low kinetic
energies (10 - 100 meV), inert He-atom beams probe the topmost layer
of the surface in a non-destructive manner. In addition, momenta of
thermal-energy He atoms are of the order of the surface phonon
momenta, making HAS suitable for scanning the entire surface Brillouin
zone. Therefore helium atom scattering is our method of choice for the
study of the structure and vibrations of the surface.

A mono-energetic supersonic beam of thermal-energy neutral He atoms is
generated by a nozzle-skimmer assembly and directed to the scattering chamber
through various collimating slits.  The energy of the He beam can be varied
from $E_i= 25$ meV to $E_i=65$ meV by changing the nozzle temperature from
110 K to 300 K.  The temperature is controlled and measured by a diode sensor
together with a temperature controller (Scientific Instruments, Model
9600-1). Polar rotation of the sample allows the variation of the He beam
incident angle $\theta_i$ with respect to the surface normal. The scattered
He beam is collected by an angle resolved detector mounted on a two-axis
goniometer which allows the scattering angle $\theta_f$ to be varied
independently. The detector is comprised of an electron gun and a
multichannel plate (MCP) electron multiplier. The electron gun generates a
well-collimated, monoenergetic electron beam crossing the He beam at right
angles. The energy of the electron beam is tuned to excite the He atoms to
their first excited metastable state (2$^3S$ He$^*$) upon
impact. Deexcitation of a He$^*$ atom at the surface of the MCP leads to the
ejection of an electron which generates an electron cascade that is then
collected by the anode of the multiplier. By electronically pulsing the
electron gun a gate function is created for time-of-flight (TOF) measurements
in the inelastic HAS mode. The details of the detection scheme are given in
Ref \onlinecite{Martini87}. All the measurements are performed with the
sample surface at room temperature.

By writing the He-atom wave vector as ${\bm k}= ({\bm K}, k_z)$, where
${\bm K}$ is the component parallel to the surface, conservation of
momentum and energy for in-plane He scattering can be expressed as
\begin{equation}
\Delta {\bm K} = k_f\sin\theta_f - k_i \sin\theta_i = {\bm G}+ {\bm
Q},
\label{k-conserv}
\end{equation} 
and
\begin{equation}
\Delta E = \hbar \omega({\bm Q}) = E_f -E_i
=\frac{\hbar^2}{2m}\,(k_f^2  - k_i^2),
\label{e-conserv}
\end{equation} 
where subscripts $i$ and $f$ denote incident and scattered beams,
respectively and $\Delta {\bm K}$ is the momentum transfer parallel to
the surface. ${\bm G}$ is a surface reciprocal lattice vector, ${\bm
Q}$ is the surface phonon wave vector and $\hbar \omega({\bm Q})$ is
the corrsponding surface phonon energy. By eliminating $k_f$ from the
above equations one obtains the so-called {\em scan curve} relations
which are the locus of all the allowed $\Delta E$ and $\Delta {\bm K}$
as dictated by the conservation relations:
\begin{equation}
 \Delta E = E_i\bigg[\bigg(\frac{\sin\theta_i + \Delta
 K/k_i}{\sin\theta_f}\bigg)^2 - 1 \bigg]. 
\label{scan}
\end{equation}
The intersections of these scan curves with the phonon dispersion
curves define the kinematically allowed inelastic events for a fixed
geometric arrangement. Thus, by systematically changing $E_i$,
$\theta_i$ and $\theta_f$ the entire dispersion curves can be
constructed.

\section{RESULTS}
\subsection{Elastic He Scattering}
\begin{figure}
\centering 
\begin{tabular}{c}
  \begin{minipage}{3.25in}
  \psfrag{x1}{$E_i = 26$ meV}
  \psfrag{x2}{$\theta_i = 40^\circ$}
  \psfrag{x3}[c]{$\Delta K$ ({\AA}$^{-1}$)}
  \psfrag{x4}[c]{Intensity (arb. units)}
  \psfrag{x5}[c][][0.8]{($\bar{2}0$)}
  \psfrag{x6}[c][][0.8]{($\bar{1}0$)}
  \psfrag{x7}[c][][0.8]{(00)}
  \psfrag{x8}[c][][0.8]{(10)}
  \psfrag{x9}{$\langle 100 \rangle$} 
  \includegraphics*[width=3.25in]{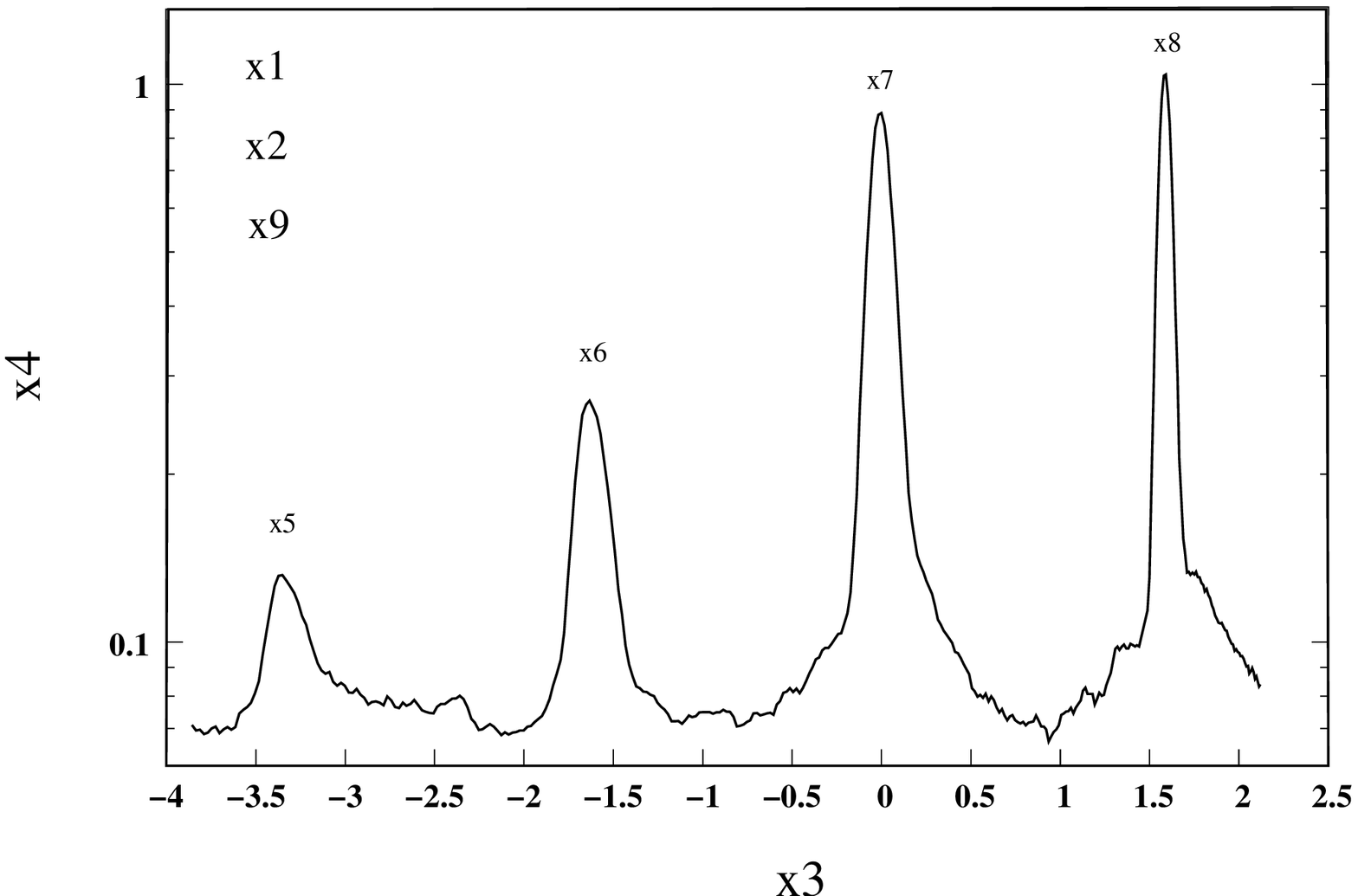}
 \end{minipage} \\
\\
\\
 \begin{minipage}{3.2in}
  \psfrag{x1}{$E_i = 25$ meV}
  \psfrag{x2}{$\theta_i = 35^\circ$}
  \psfrag{x3}[c]{$\Delta K$ ({\AA}$^{-1}$)}
  \psfrag{x4}[c]{Intensity (arb. units)}
  \psfrag{x5}[c][][0.8]{($\bar{1}\bar{1}$)}
  \psfrag{x6}[c][][0.8]{(00)}
  \psfrag{x7}[c][][0.8]{(11)}
  \psfrag{x8}{$\langle 110 \rangle$}
  \includegraphics*[width=3.2in]{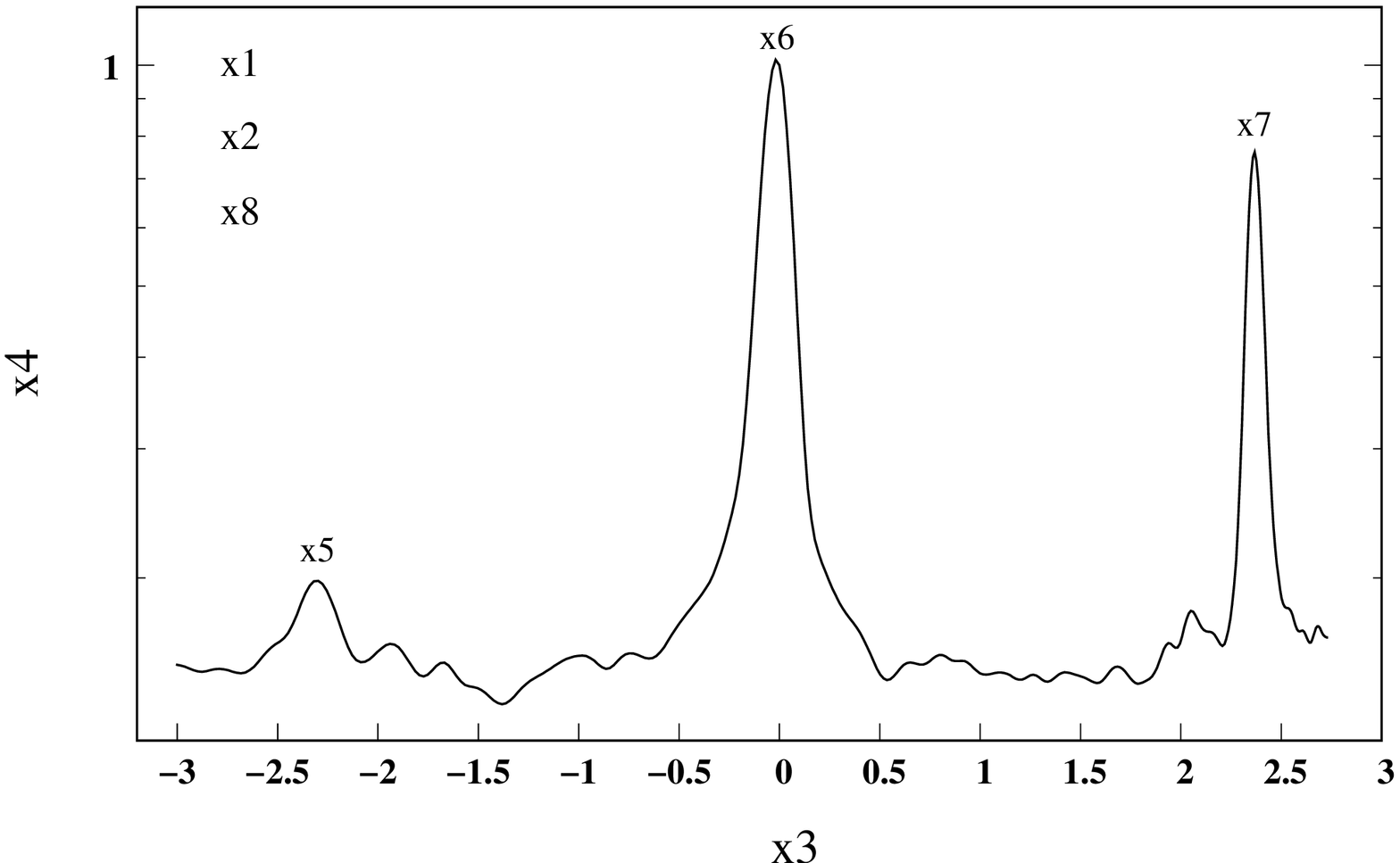}
 \end{minipage}
\end{tabular}
\caption{Typical diffraction patterns along $\langle 100 \rangle$ and $\langle 110
  \rangle$ azimuth directions for cleaved Sr$_2$CuO$_2$Cl$_2$(001)
  surface. The intensity scale is logarithmic.}
\label{diff}
\end{figure}
Figure \ref{diff} shows two typical He diffraction patterns from the
(001) surface of Sr$_2$CuO$_2$Cl$_2$ along $\langle 100 \rangle$ and
$\langle 110 \rangle$ symmetry directions for $\theta_i=40^\circ$ and
$35^\circ$, respectively.  The curves are shown as a function of the
parallel momentum transfer $\Delta{\bm K}$.  The surface periodicity
is obtained from the positions of the diffraction peaks along each
direction resulting from the Bragg relation
\begin{equation}
\Delta K = n\,G_{\langle 100 \rangle}, \;\;\;\;\;\; {\rm or} \;\;\;\;\;\; 
\Delta K = n\,G_{\langle 110 \rangle}, 
\label{Bragg}
\end{equation}
where integer $n$ indicates the order of diffraction. From these and
the position of Bragg peaks obtained from other diffraction patterns
taken at various incident angles, the value of $a$ = 3.97 {\AA} for
the surface lattice parameter was obtained.  This value is very close
to the lattice parameter reported for the bulk \cite{Miller90}. This,
and the fact that no systematic satellite peaks in the diffraction
patterns were observed, suggest a bulk like termination at the surface and
indicates surface stability against reconstruction.

\begin{figure}
\begin{center}
  \includegraphics*[width=3in]{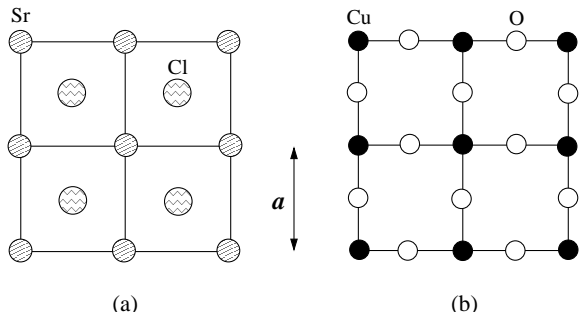}
  \caption{Schematic illustration of two different possible (001) surfaces created
    by cleaving Sr$_2$CuO$_2$Cl$_2$: (a) SrCl and (b) CuO$_2$.}
\label{surf-structure}
\end{center}
\end{figure}

In general, cleaving the Sr$_2$CuO$_2$Cl$_2$ can result in two
different surfaces with different compositions as shown in Figure
\ref{surf-structure}. SrCl termination of the crystal results in a
non-polar surface which, according to Tasker's classification of ionic
surfaces \cite{Tasker79}, consists of a neutral repeat unit of SrCl -
CuO$_2$ - SrCl, and is stable. However, termination at a CuO$_2$
layer, results in a repeat unit of CuO$_2$ - SrCl which has a
non-vanishing dipole moment perpendicular to the surface. Such a polar
surface cannot be stable because of the divergence of the
electrostatic energy contribution to the surface energy. Hence, such
termination would be allowed only if the surface undergoes a
reconstruction resulting in cancellation of the dipole moment
perpendicular to the surface. The apparent surface stability against
reconstruction on the (001) surface of Sr$_2$CuO$_2$Cl$_2$ suggests
that the top-most layer is indeed SrCl and {\em not} CuO$_2$. This is
also supported by previous LEED and x-ray photoemission spectroscopy
studies of the Sr$_2$CuO$_2$Cl$_2$(001) surface
\cite{Durr00,Boske97,Koitzsch02} which suggest that SrCl is the
top-most layer.

From the intensity analysis of the diffraction peaks, one can obtain
the surface corrugation function. A two dimensional surface
corrugation consistent with the $4mm$ symmetry of the surface was
chosen as
\begin{align}
\zeta(x,y) = & \, \zeta_1 \,\bigg[ \cos\bigg(\frac{2\pi x}{a}\bigg) + \cos\bigg(\frac{2\pi
y}{a}\bigg) \bigg]  \notag \\
+ &\,  \zeta_2 \,\bigg[ \cos\bigg(\frac{2\pi (x+y)}{a}\bigg) + \cos\bigg(\frac{2\pi
(x-y)}{a}\bigg) \bigg],
\label{corrugation} 
\end{align} 
where $\zeta_1$ and $\zeta_2$ denote the corrugation amplitudes and are taken
as model parameters. Values of $\zeta_1 = 0.01$ {\AA} and $\zeta_2 = 0.15$
{\AA} were obtained by fitting Bragg peak intensities calculated within the
framework of the eikonal approximation \cite{Farias98} and using an
attractive potential well depth of $D = 10$ meV to the experimental
intensities. To back up this calculation we also obtained the corrugation
amplitudes by calculating the surface charge density $\rho(x,y,z)$, since the
corrugation function is the locus of the classical turning points for He
atoms at a fixed surface charge density contour. $\rho(x,y,z)$ was
constructed as a superposition of the individual atomic charge densities at a
SrCl surface. The atomic charge densities for Sr and Cl were derived using
the atomic wave functions of Herman and Skillman \cite{Herman63}.  The values
of $\zeta_1$ and $\zeta_2$ calculated at a constant contour of $\rho(x,y,z)$
at a turning point of 2.7 {\AA} above the surface were 0.01 {\AA} and 0.147
{\AA}, which are very close to the values obtained by the eikonal
approximation.

\subsection{Surface Phonons}
\begin{figure}
\begin{center}
  \includegraphics*[width=2in]{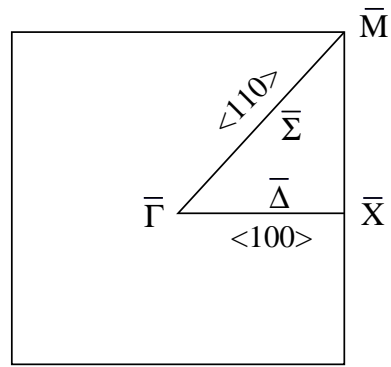}
  \caption{Surface Brillouin zone and the high symmetry directions.}
\label{SBZ}
\end{center}
\end{figure}

\begin{figure}
\centering 
\begin{tabular}{c}
  \begin{minipage}{3.3in}
  \psfrag{x1}[c]{TOF ($\mu$ sec)}
  \psfrag{x2}[c]{Intensity (arb. units)}
  \psfrag{x3}{$E_i = 31.6$ meV}
  \psfrag{x4}{$\theta_i = 44^\circ$}
  \psfrag{x5}{$\theta_f = 53^\circ$}
  \psfrag{x6}[c][][0.8]{Elastic}
  \psfrag{x7}[c]{$\langle 100 \rangle$}
  \includegraphics*[height=3.2in,width=3.1in]{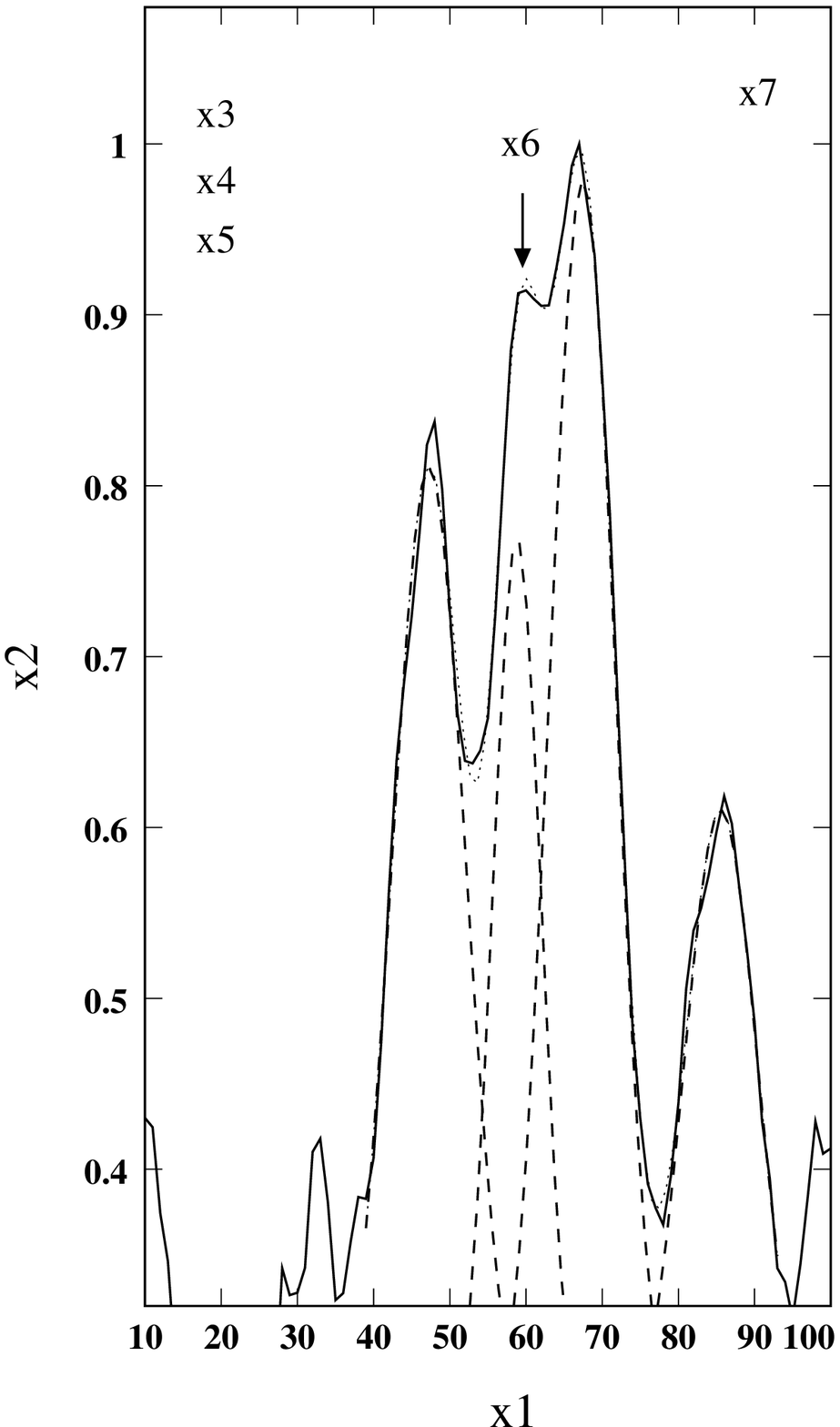}
 \end{minipage} \\
\\
\\
 \begin{minipage}{3.3in}
  \psfrag{x1}[c]{TOF ($\mu$ sec)}
  \psfrag{x2}[c]{Intensity (arb. units)}
  \psfrag{x3}{$E_i = 26.4$ meV}
  \psfrag{x4}{$\theta_i = 41.5^\circ$}
  \psfrag{x5}{$\theta_f = 38.5^\circ$}
  \psfrag{x6}[c][][0.8]{Elastic}
  \psfrag{x7}[c]{$\langle 110 \rangle$}
  \includegraphics*[height=3.2in,width=3.1in]{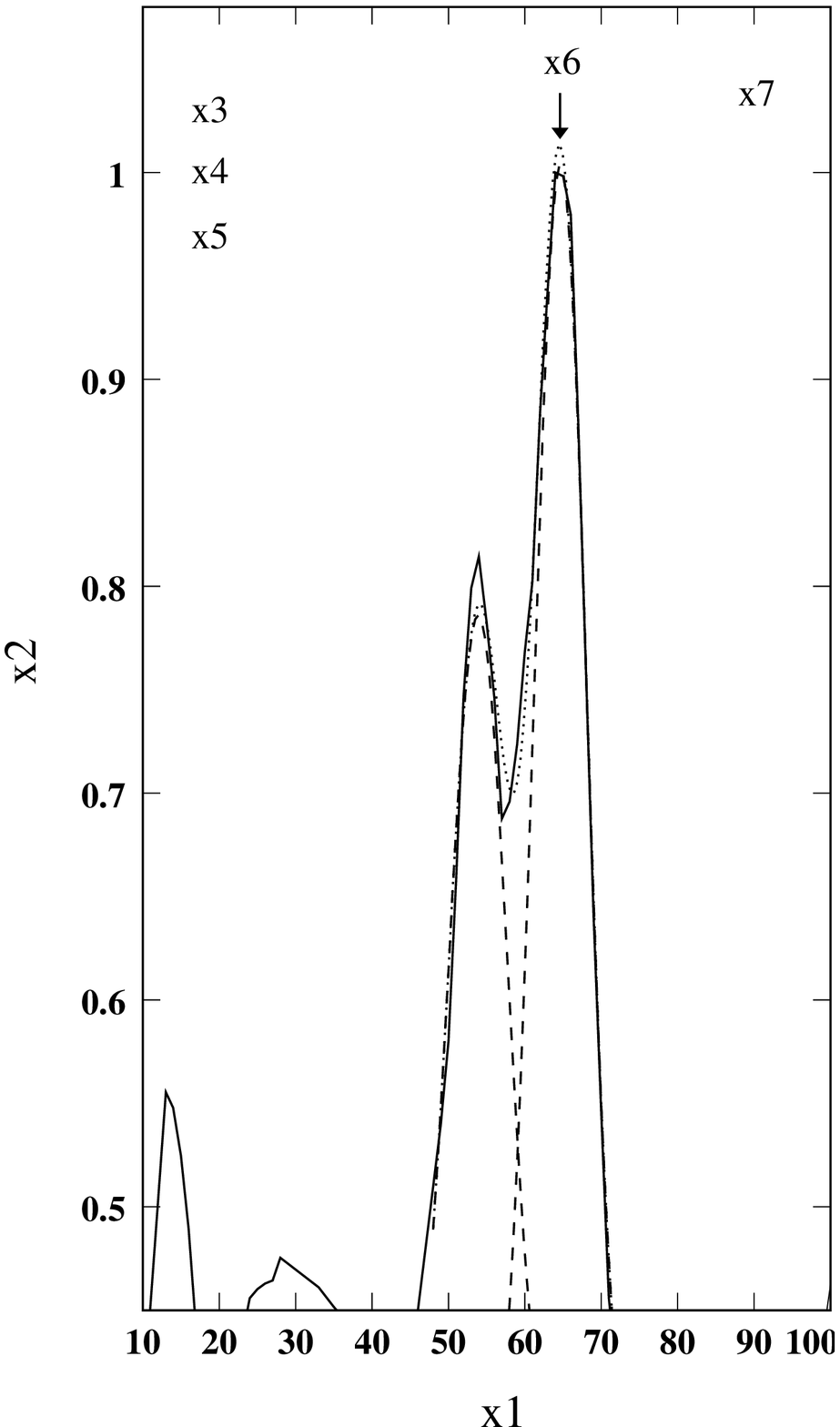}
 \end{minipage}
\end{tabular}
\caption{Two typical TOF spectra along $\langle 100 \rangle$ and $\langle 110
  \rangle$ directions (solid lines). Dashed and dotted lines are the Gaussian
  fits to the peaks.}
\label{tof}
\end{figure}

Surface phonon energies were measured along the $\bar{\Delta}$ and
$\bar{\Sigma}$ directions of the surface Brillouin zone (SBZ), shown
in Figure \ref{SBZ}.
Two typical TOF spectra showing diffusive elastic, as well as
inelastic peaks are shown in Figure \ref{tof}. The data were collected
using beam energies in the range of 25 - 65 meV.  The high energy He
beam ($E_i\sim 65$ meV) was used to probe the high-frequency surface
phonon modes. The energy resolution for the lowest energy beam was
about 3 meV. Phonon energies and momenta were calculated from the
TOF peak positions, after a Gaussian fit to the peaks, using
Eqs.(\ref{k-conserv}) and (\ref{e-conserv}). To characterize the
ensuing phonon dispersion points lattice dynamical analysis for the
bulk and surface (slab calculation) were carried out as outlined in
the next section.

\section{LATTICE DYNAMICAL ANALYSIS}
\subsection{Static Equilibrium and Bulk Lattice Dynamics}
In lattice dynamical studies of the cuprates and HTSC parent compounds, the
construction of a dynamical matrix based on pair potentials rather than mere
force constants is far more advantageous in the sense that the underlying
physics is transparent and many of the potential parameters for similar pair
interactions can be transferred from one compound to another
\cite{Rampf93,Chaplot95}. This is specially useful for compounds like
Sr$_2$CuO$_2$Cl$_2$ for which experimental phonon frequencies are limited to
four infrared (IR)-active phonon modes at the center of the Brillouin zone
(BZ) \cite{Tajima91,Zibold96}, and no neutron scattering data for phonons are
available. However, experimental and theoretical studies of bulk lattice
dynamics of the isostructural La$_2$CuO$_4$ are extensive and contain
detailed information about relevant model potentials
\cite{Mostoller90,Collins89,Eklund89,Weber88,Boni88,Birgeneau87}. Therefore,
for some of the pair interactions which are similar in both compounds (such
as Cu-O and O-O), the existing potential parameters for La$_2$CuO$_4$ can be
transferred to Sr$_2$CuO$_2$Cl$_2$. In Sr$_2$CuO$_2$Cl$_2$, with seven atoms
per primitive cell, more model parameters can be determined by fulfilling the
static equilibrium conditions which state that the forces on the particles in
their equilibrium positions should vanish. This treatment also ensures the
consistency of the static and dynamical properties of the crystal. It should
be noted, however, that satisfying static equilibrium conditions, does not
guarantee any dynamical stability, namely ensuring the reality of the phonon
frequencies throughout the BZ.

The model used for both static equilibrium calculations and lattice dynamics
(shell-model) of bulk Sr$_2$CuO$_2$Cl$_2$, incorporates two-body central
potentials, namely Coulomb potential $V_{ij}^{\rm C}(r) = Z_i\,Z_j\,e^2/r$
for the long-range interactions and either Born-Mayer (BM) $V_{ij}^{\rm
BM}(r) = a_{ij}\,e^{-b_{ij}r}$ or Buckingham $V_{ij}^{\rm B}(r) =
a_{ij}\,e^{-b_{ij}r}-c_{ij}/r^6$ potentials for the short-range
interactions. Short-range interactions were limited to nearest neighbor pairs
for Cu-O, O-O, Sr-O, Cl-Cl, Cu-Cl, and O-Cl, and to nearest and next nearest
neighbors for Sr-Cl. Potential parameters were obtained from existing
literature for Cu-O, O-O, Sr-O and Cl-Cl. Static equilibrium conditions
provided four equations and were based on the methods developed in
Refs. \onlinecite{Rampf93} and \onlinecite{Boyer73}.

Table \ref{parameters} lists all potential parameters obtained either
from the literature or by satisfying equilibrium conditions.
\begin{table}
\caption{Potential parameters for pair interactions in 
         Sr$_2$CuO$_2$Cl$_2$ and shell-model parameters.}
\begin{ruledtabular}
\begin{tabular}{cccc}
       & \multicolumn{3}{c}{Potential Parameters} \\
Ions   & $a$ (eV) & $b$ (\AA$^{-1}$) & $c$ (eV \AA$^6$)\\\hline
Cu-O\footnotemark[1]& 5814.375 & 4.762  & 0 \\
O-O\footnotemark[1] & 1146.25  & 3.279  & 0 \\
Sr-O   & 1950    & 2.978\footnotemark[2]& 0 \\
Cl-Cl\footnotemark[3]& 18498.3  & 3.65 & 99.87\\
Sr-Cl  & 62037.2  & 4.064 & 0  \\
Cu-Cl  & 103.55   & 3.0     & 0  \\
O-Cl   & 1161.75  & 3.0     & 0  \\ \hline
       & \multicolumn{3}{c}{Shell-Model Parameters} \\ 
Ion    &  $Z$ ($e$) & $Y$ ($e$)  & $K$ (eV/ \AA$^2$) \\ \hline
Cu     & 2          & 2.6        & 40  \\
Sr     & 2          & 1.9        & 30  \\
O      & -2         & -3.1       & 24  \\ 
Cl     & -2         & -1.8       & 5   \\
\end{tabular}
\label{parameters}
\end{ruledtabular}
\footnotetext[1]{Ref. \onlinecite{Mostoller90}}.
\footnotetext[2]{Ref. \onlinecite{Kovaleva04}}.
\footnotetext[3]{Ref. \onlinecite{Venkataraman}}
\end{table}
These parameters were then incorporated into a shell-model to
calculate the phonon dispersion curves of bulk Sr$_2$CuO$_2$Cl$_2$. In
order to ensure dynamical stability over the entire BZ, $a_{\rm Cu
- O}$ was increased by \%15.

The shell-model parameters, which include the ionic charge $Z$, the
shell charge $Y$ and the intra-ion shell-core force constant $K$, are
also listed in Table \ref{parameters}. These parameters were adjusted
to give a best fit to the four measured \cite{Tajima91,Zibold96}
IR-active phonon frequencies at the $\Gamma$-point. The agreement
between measured and calculated frequencies is reasonably good, 
as can be seen in Table \ref{gamma}, and the discrepancy between
measurement and calculation is at most \%5.
\begin{table}
\caption{Comparison of the four measured infrared-active frequencies
          for Sr$_2$CuO$_2$Cl$_2$ at the center of the Brillouin zone
          ($q=0$) with the values calculated by the shell model in this
          work. All the frequencies are in units of cm$^{-1}$.}
\begin{ruledtabular}
\begin{tabular}{cccc}
 & \multicolumn{2}{c}{Measured Frequencies} & \\
Mode & Ref. \onlinecite{Tajima91} & Ref. \onlinecite{Zibold96} & Calculated Frequencies\\\hline
$E_u(1)$  & 525       &           512       &     522.41      \\
$E_u(2)$  & 351       &           339       &     334.37      \\
$E_u(3)$  & 176       &           173       &     182.67      \\ 
$E_u(4)$  & 140       &           138       &     143.51      \\
\end{tabular}
\label{gamma}
\end{ruledtabular}
\end{table}

\begin{figure}
\begin{center}
  \includegraphics*[width=3.5in]{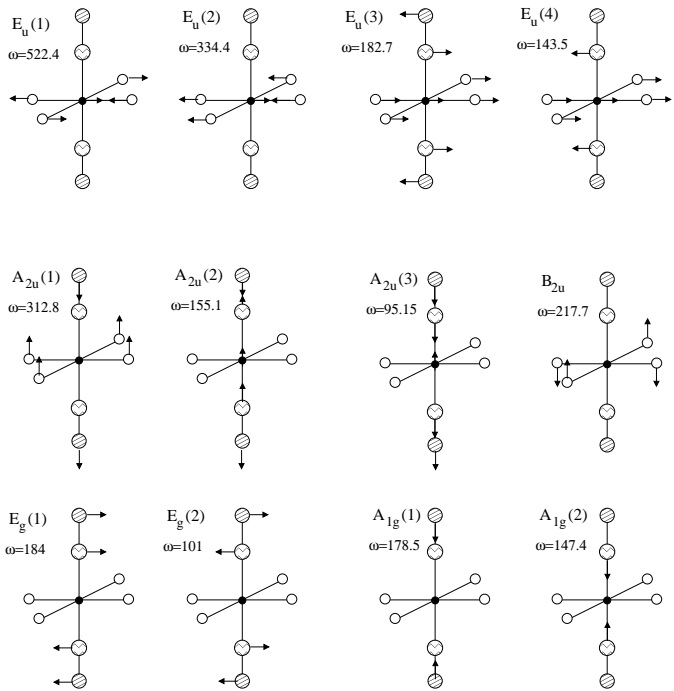}
  \caption{The eigenvectors and frequencies of the transverse optic (TO) phonon modes of
Sr$_2$CuO$_2$Cl$_2$ at ${\bm q}=0$ as obtained from the shell-model. All the
frequencies are in units of cm$^{-1}$.}
\label{gamma-modes}
\end{center}
\end{figure}

The eigenvectors and frequencies obtained from the shell-model for the
transverse optic (TO) phonon modes of Sr$_2$CuO$_2$Cl$_2$ at ${\bm
q}=0$ are depicted in Figure \ref{gamma-modes}.
The four $E_u$ modes are the IR-active modes listed in Table
\ref{gamma} and are doubly degenerate. The atomic displacements
corresponding to these modes are in agreement with mode
assignments of Ref. \onlinecite{Tajima91}. $E_u(1)$ is a Cu-O
stretching mode, $E_u(2)$ is a Cu-O bending mode, $E_u(3)$ is the
translational vibration of a Sr-atom layer against the octahedron and
$E_u(4)$ is the apical bending mode of Cl against the Cu-O unit
\cite{Tajima91,Zibold96}.  The non-degenerate $A_{2u}$ modes are also
IR-active and correspond to the atomic displacements along
$z$-direction. The doubly-degenerate $E_g$ and non-degenerate $A_{1g}$
modes are Raman-active and $B_{2u}$ is silent.
  
\begin{figure}
\begin{center}
  \includegraphics*[width=3in]{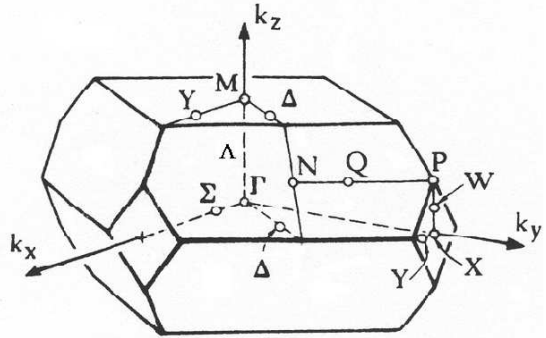}
  \caption{The bulk BZ for body-centered tetragonal structure with the
    labelling of symmetry directions.}
\label{bulk-bz}
\end{center}
\end{figure}

\begin{figure*}
   \includegraphics*[width=6in]{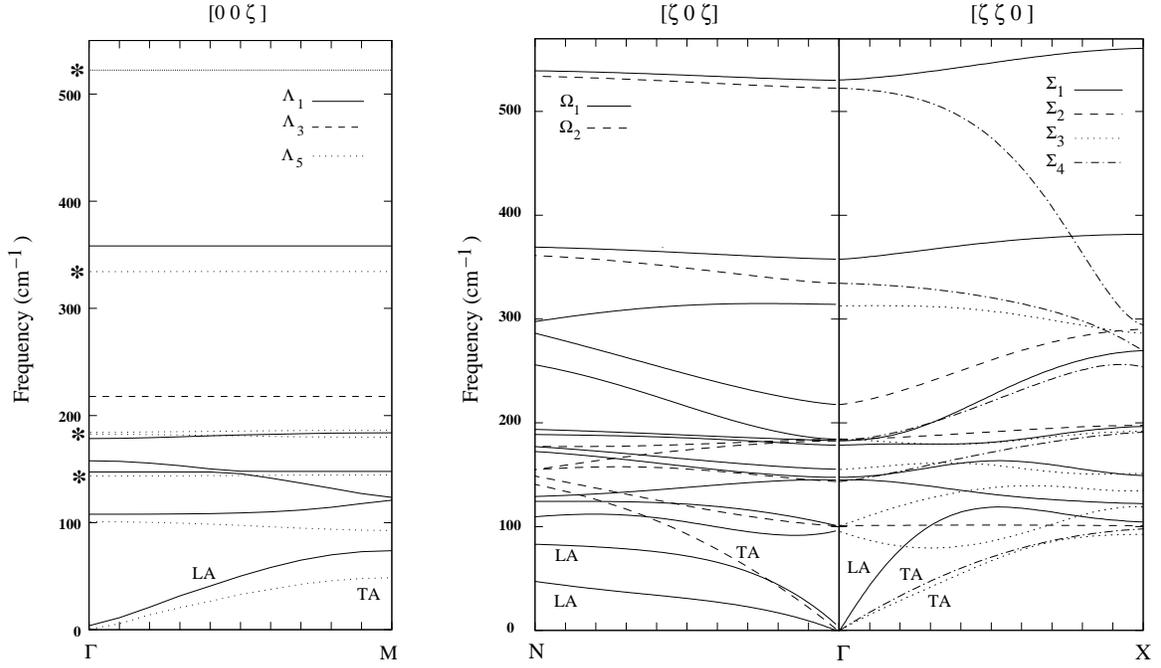}
\caption{Phonon dispersion curves along three symmetry directions in the bulk BZ
  calculated by the shell-model of Table \ref{parameters}. The four measured
  \cite{Tajima91,Zibold96} IR-active $E_u$ frequencies at $\Gamma$ point are
  labeled by asterisks.}
\label{bulk-disp}
\end{figure*}

The bulk BZ for body-centered tetragonal structure is shown in Figure
\ref{bulk-bz} with the labelling of high-symmetry points and
directions. Bulk phonon dispersion curves, calculated along
$\Lambda$, $\Sigma$ and $\Gamma$N directions using the shell-model of
Table \ref{parameters}, are shown in Figure \ref{bulk-disp}. The
longitudinal acoustic (LA) and transverse acoustic (TA) modes along
all three directions are also labeled. It can be inferred form these
dispersion curves that the dynamics of the Sr$_2$CuO$_2$Cl$_2$ bulk is
stable throughout the BZ without any imaginary frequency.

\subsection{Slab Lattice Dynamics}
\begin{figure}
\begin{center}
  \includegraphics*[width=3in]{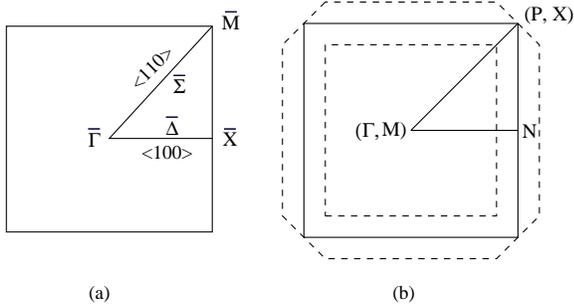}
  \caption{(a) SBZ and (b) projection of the bulk BZ on the surface (dashed-lines).}
\label{projection}
\end{center}
\end{figure}

Our studies of the surface lattice dynamics of Sr$_2$CuO$_2$Cl$_2$
were also based on the shell-model. In these studies we employed slab
geometries comprised of SrCl - CuO$_2$ - SrCl repeat unit with 21 and
33 layers and with periodic boundary conditions in the plane of the
slab.  The slab was bounded by two SrCl surface layers. The total
number of degrees of freedom in the 33-layer slab is 231. The
projection of the bulk BZ on the surface is illustrated in Figure
\ref{projection}.  Projections of all the bulk phonon modes that lie
inside the SBZ contribute to the bulk bands in slab dispersion curves.

\begin{figure*}
\psfrag{x1}[][][1.2]{$\bar{{\rm M}}$}
\psfrag{x2}[][][1.2]{$\bar{\Gamma}$}
\psfrag{x3}[][][1.2]{$\bar{{\rm X}}$}
\psfrag{x4}[][][1.2]{$\bar{\Sigma}$}
\psfrag{x5}[][][1.2]{$\bar{\Delta}$}
\psfrag{x6}[c][][1.2]{$\langle$110$\rangle$}
\psfrag{x7}[c][][1.2]{$\langle$100$\rangle$}
\psfrag{x8}[c][][1.2]{Frequency (cm$^{-1}$)}
  \includegraphics*[width=6in]{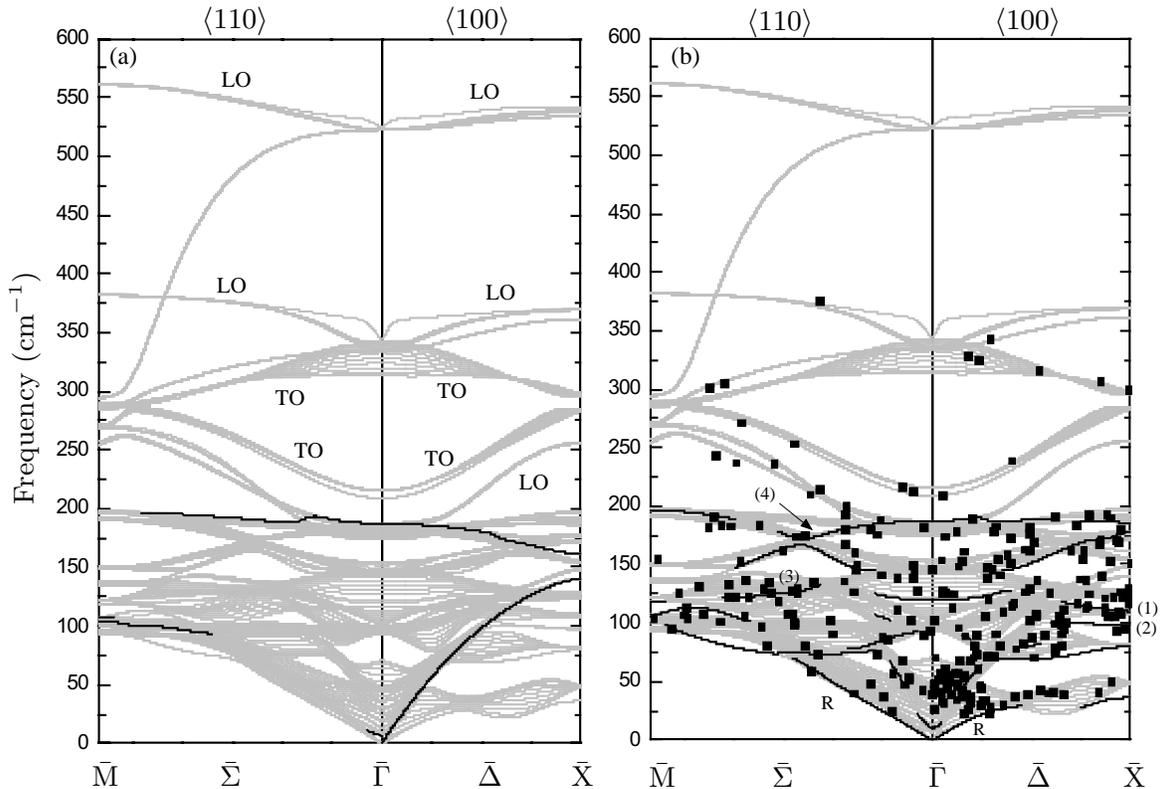}
\caption{Results of the slab calculation for a 33-layer slab of
Sr$_2$CuO$_2$Cl$_2$ with SrCl surface termination along $\langle$100$\rangle$ and
$\langle$110$\rangle$ directions. The grey curves show the bulk bands.
In (a) black solid lines correspond to surface modes with SH polarization,
while in (b) the black lines show SP surface modes.  
The filled squares are the experimental data obtained by inelastic He
scattering. Modes labeled by (1), (2), (3) and (4) are discussed in the text.
The acoustic Rayleigh modes are labeled by R.}
\label{slab-disp}
\end{figure*}

Figure \ref{slab-disp} shows the results of the slab calculations
along both $\langle 100\rangle$ and $\langle 110\rangle$ directions of
the SBZ where unmodified bulk shell-model parameters (Table
\ref{parameters}) were used throughout the slab. The grey areas and
curves in Figures \ref{slab-disp}a and \ref{slab-disp}b are
projections of the bulk modes on the SBZ (bulk bands). In Figure
\ref{slab-disp}a solid black lines identify dispersion curves for
shear horizontal (SH) surface phonon polarizations. These modes are
polarized perpendicular to the sagittal plane (defined by the surface
normal and the phonon wave vector) and parallel to the surface; they are
odd with respect to reflection in the sagittal plane. Because in our
scattering geometry the sagittal and scattering planes
coincide and contain a high symmetry surface direction, SH modes would not be
detected. In Figure \ref{slab-disp}b, solid black lines correspond
to surface modes which are polarized in the sagittal plane (SP modes)
and the experimental points are shown by filled squares.

From the surface dispersion curves of Figure \ref{slab-disp} it is
clear that all the SrCl surface modes are below 200 cm$^{-1}$. The
high energy bulk bands, labeled as longitudinal optic (LO) and
transverse optic (TO) bands in Figure \ref{slab-disp}a, correspond to
the vibrations of Cu and O atoms in the bulk layers, consistent with
the results from bulk dispersion curves.

Below 200 cm$^{-1}$, the agreement between the experimental points and
the surface modes is reasonably good. In both directions there are
experimental points which clearly fall along the surface SP modes. One
example is the SP mode around 115 cm$^{-1}$ along
$\langle$100$\rangle$ (labeled by (1) in Figure \ref{slab-disp}b),
which corresponds to vibration of surface Sr atom along $z$ direction
and Cl atom along $x$ direction. Another example is the mode labeled
by (2) along $\langle$100$\rangle$ direction which corresponds to the
displacement of the surface Cl atom along $z$. Another area where
there is agreement between theory and experiment is the region with a
high density of bulk bands from $\sim$ 30 to 70 cm$^{-1}$, close to
the $\bar{\Gamma}$ point along $\langle$100$\rangle$.  Because of the
high density of bulk states in this region, a large number of
experimental points have also been detected.  Along
$\langle$110$\rangle$ direction, mode (3) is associated with the
vibrations of the surface Sr atom along $z$ and mode (4) corresponds
to displacements of both Sr and Cl surface atoms in the $xy$ plane.

Along both directions, the acoustic Rayleigh modes are labeled by
R. The scarcity of experimental points along the Rayleigh modes is
attributed to the limitation of the energy resolution that has
prevented modes with energies lower than $\sim 25$ cm$^{-1}$ to be
clearly identified.

Because of the systematic agreement between the SrCl surface modes and
experimental dispersion points, no further adjustments of model parameters
were necessary. 

\begin{figure*}
\psfrag{x1}[][][1.2]{$\bar{\Gamma}$}
\psfrag{x2}[][][1.2]{$\bar{{\rm X}}$}
\psfrag{x3}[][][1.2]{$\bar{\Delta}$}
\psfrag{x4}[c][][1.2]{$\langle$100$\rangle$}
\psfrag{x5}[c][][1.2]{Frequency (cm$^{-1}$)}
 \includegraphics*[width=6in]{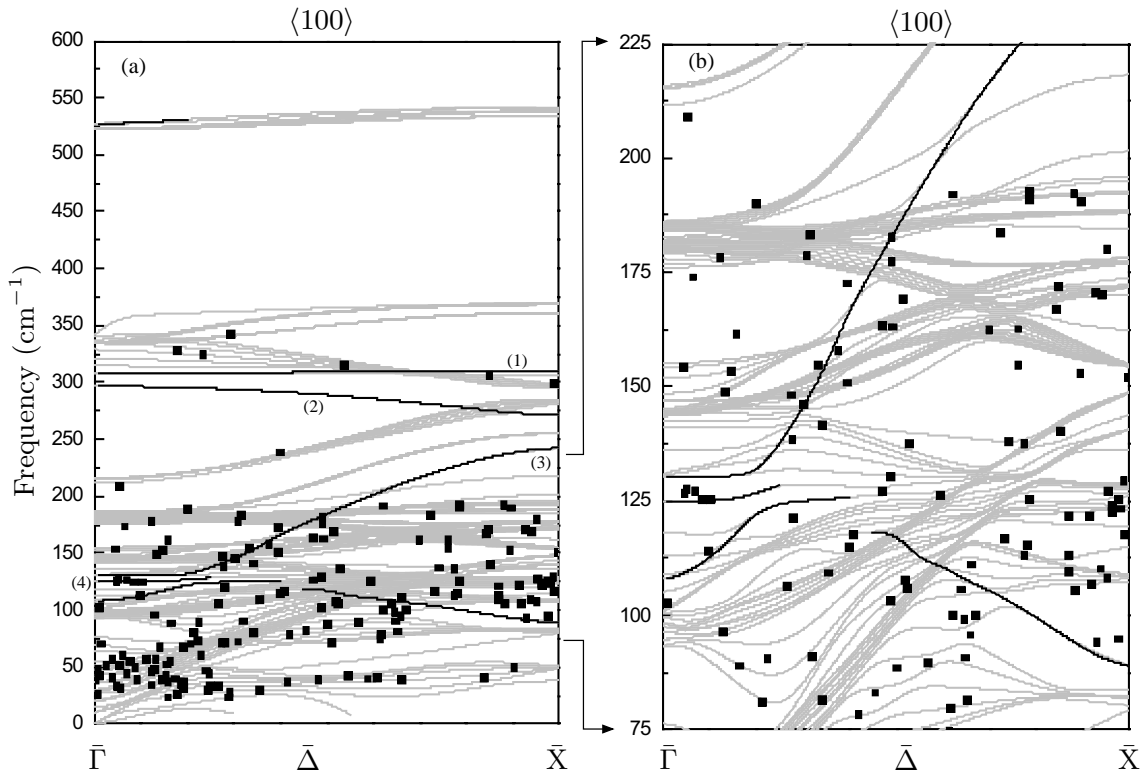}
\caption{Results of the slab calculation for a 21-layer slab of
Sr$_2$CuO$_2$Cl$_2$ with CuO$_2$ surface termination along $\langle$100$\rangle$
direction. The grey curves are bulk bands, the black solid lines are
the surface modes with the SP symmetry and the filled squares are the
experimental data. Modes labeled by (1), (2), (3) and (4) are discussed in
the text. (b) is the expansion of (a) from 75 to 225 cm$^{-1}$.}
\label{slab-cuo}
\end{figure*}

In order to confirm that the cleaved (001) surface of Sr$_2$CuO$_2$Cl$_2$
consists of only SrCl layers and not CuO$_2$, we implemented the slab
calculations for a 21-layer slab with a CuO$_2$ surface termination. In order
to maintain the charge neutrality, the slab had to be made asymmetric,
i.e. the other surface of the slab was a SrCl layer. Employing bulk
shell-model parameters resulted in instabilities of several modes away from
$\bar{\Gamma}$. Efforts to completely suppress these instabilities were not
successful. However, we found that setting $b_{{\rm  Cu-O}}$ in the surface
layer to 5.2 {\AA}$^{-1}$ resulted in confining the instabilities to the
lowest lying modes. In these calculations the optic modes of the
CuO$_2$ surface were stable as shown in Figure \ref{slab-cuo} for SP surface
modes along $\langle$100$\rangle$ direction. The instability of the lowest
lying acoustic modes away from the $\bar{\Gamma}$ point is clear in Figure
\ref{slab-cuo}a.

By contrast to the results of slab calculation for SrCl surface, we did not
find systematic agreement between the experimental data and calculated CuO$_2$
surface modes. For example no experimental points exist close to the surface modes
labeled (1), (2) and (3) in Figure \ref{slab-cuo}. Mode (1)
corresponds to the vibration of O(2) surface atoms in the $x$
direction, mode (2) corresponds to the displacements of the surface
O(1) and O(2) atoms along $z$ and mode (3) is associated with
vibrations of surface Cu and O(2) atoms along $z$ and O(1) atoms in
the $xz$ plane. The coincidental agreement of a few points with some CuO$_2$
surface modes (e.g. mode (4) around 125 cm$^{-1}$), is more accidental than
systematic.     

\section{CONCLUSION}
Cleaved (001) surfaces of the layered perovskite Sr$_2$CuO$_2$Cl$_2$
were prepared in UHV, and the structure and phonon dispersions of the
surface were studied experimentally using elastic and inelastic HAS along
high-symmetry directions of the SBZ. The position of the Bragg peaks
in the diffraction patterns, obtained for different incident angles,
revealed a surface periodicity consistent with bulk termination, 
without any evidence of surface reconstruction. This fact, along with
previous LEED and x-ray photoemission results
\cite{Boske97,Koitzsch02,Durr00} confirms that the resulting surface is
non-polar and stable which favors a SrCl surface termination. A corrugation function
for the SrCl surface was constructed with the aid of eikonal approximation
using measured diffraction peak intensities and was in agreement with
a corrugation function obtained from surface charge density calculations.

Bulk lattice dynamical calculations based on the shell-model were
carried out by transferring some of the potential parameters from existing
models for other cuprates like La$_2$CuO$_4$ and obtaining the rest by
solving static equilibrium equations. The shell-model parameters were
adjusted to fit the four experimentally measured IR-active phonon
frequencies at ${\bm q}=0$. No instability were observed over the
entire BZ.

Surface dynamics was investigated using slab geometry and the same
shell-model. The calculations for a 33-layer slab with SrCl surfaces and a
21-layer slab with a CuO$_2$ surface revealed two important features. First
that the experimental surface phonon dispersions obtained by inelastic HAS
agreed quite reasonably and systematically with a SrCl terminated slab
model. Second, the shell-model for a CuO$_2$ terminated slab could not be
stabilized, despite all efforts. This and the fact that no experimental
dispersion points lie systematically along the CuO$_2$ surface modes is
another sign that the surface  layer is SrCl and CuO$_2$ sheets do not exist
on the surface.
\begin{acknowledgments}
The authors would like to thank J. R. Manson for helpful discussions.
This work was supported by the U.S. Department of Energy under Grant
No. DE-FG02-85ER45222 and in part by MRSEC Program of the National Science
Foundation under award number DMR 02-13282. 
\end{acknowledgments}

%\bibliographystyle{/home/maryamf/Thesis/prsty} 
%\bibliography{/home/maryamf/Thesis/main}

\end{document}